\documentclass[a4paper, aps, prl, showpacs, twocolumn, superscriptaddress, nofootinbib]{revtex4}

\usepackage{tikz}
\usepackage{xcolor}
\usepackage{dsfont}
\usepackage[utf8]{inputenc}  
\usepackage[british]{babel}  
\usepackage{graphicx} 
\usepackage{pgfplots}  
\usepackage{multirow}
\usepackage{dcolumn}   
\usepackage{bm}        
\usepackage{amssymb}   
\usepackage{amsmath}
\usepackage{mdframed}
\usepackage{theorem}

\usetikzlibrary{arrows}
\usetikzlibrary{shapes}
\usetikzlibrary{decorations.pathmorphing}
\usetikzlibrary{decorations.pathreplacing}
\usetikzlibrary{decorations.shapes}
\usetikzlibrary{decorations.text}

\DeclareMathOperator{\vrho}{\varrho}
\DeclareMathOperator{\II}{\mathcal{I}}
\DeclareMathOperator{\BB}{\mathcal{B}}

\newcommand{\bea}{\begin{eqnarray}}
\newcommand{\eea}{\end{eqnarray}}

\def\C{\hbox{$\mit I$\kern-.7em$\mit C$}}
\def\N{\hbox{$\mit I$\kern-.6em$\mit N$}}

\newcommand{\ket}[1]{\ensuremath{|#1\rangle}}

\newcommand{\bra}[1]{\ensuremath{\langle#1|}}

\newcommand{\braket}[2]{\ensuremath{\langle #1|#2\rangle}}
\newcommand{\ketbra}[1]{\ensuremath{| #1 \rangle \!\langle #1 |}}

\newcommand{\eins}{\openone}

\newcommand{\MM}{\ensuremath{\mathcal{M}}}
\renewcommand{\II}{\ensuremath{\mathcal{I}}}
\renewcommand{\vr}{\ensuremath{\varrho}}
\newcommand{\EE}{\ensuremath{\mathcal{E}}}

\newcommand{\ba}{\begin{eqnarray}}
\newcommand{\ea}{\end{eqnarray}}
\newcommand{\ban}{\begin{eqnarray*}}
\newcommand{\ean}{\end{eqnarray*}}
\newcommand{\be}{\begin{equation}}
\newcommand{\ee}{\end{equation}}

\newcommand{\tr}{\ensuremath{{\rm tr}}}
\newcommand{\dyad}[1]{\ketbra{1}}

\begin{document}

\title{Structure of temporal correlations of a qubit} 

\author{Jannik Hoffmann}
\thanks{These two authors contributed equally.}
\affiliation{Naturwissenschaftlich-Technische 
Fakultät, Universität Siegen, Walter-Flex-Straße 3, 57068 Siegen, Germany}
\author{Cornelia Spee}
\thanks{These two authors contributed equally.}
\affiliation{Naturwissenschaftlich-Technische 
Fakultät, Universität Siegen, Walter-Flex-Straße 3, 57068 Siegen, Germany}
\author{Otfried G\"uhne}
\affiliation{Naturwissenschaftlich-Technische 
Fakultät, Universität Siegen, Walter-Flex-Straße 3, 57068 Siegen, Germany}
\author{Costantino Budroni}
\affiliation{Naturwissenschaftlich-Technische 
Fakultät, Universität Siegen, Walter-Flex-Straße 3, 57068 Siegen, Germany}
\affiliation{Institute for Quantum Optics and Quantum Information (IQOQI), Boltzmanngasse 3, 1090 Vienna, Austria}

\date{\today}             

\begin{abstract}
In quantum mechanics, spatial correlations arising from measurements at 
separated particles are well studied. This is not the case, however, for 
the temporal correlations arising from a single quantum system subjected 
to a sequence of generalized measurements. We first characterize the 
polytope of temporal quantum correlations coming from the most general 
measurements. We then show that if the dimension of the quantum system 
is bounded, only a subset of the most general correlations can be realized 
and identify the correlations in the simplest scenario that can not be 
reached by two-dimensional systems. This leads to a temporal inequality 
for a dimension test, and we discuss a possible 
implementation using nitrogen-vacancy centers in diamond. 
\end{abstract}

\pacs{03.65.Ta}
\maketitle

\section{Introduction}
\begin{figure}[t!]
\begin{center}
\includegraphics[width=0.95\columnwidth]{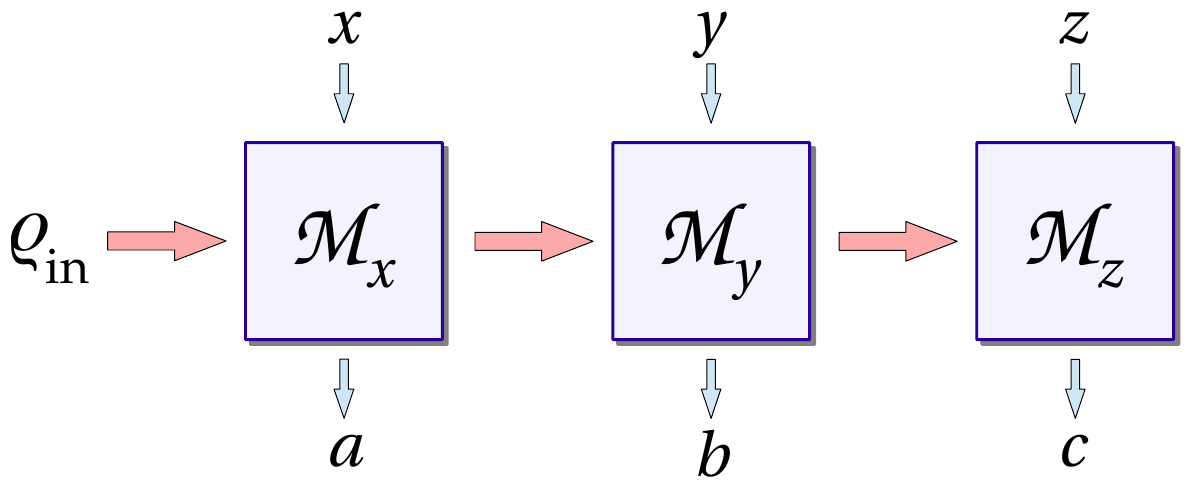} 
\end{center}
\caption{Schematic representation of the situation considered in this 
paper. A quantum system (described by a density matrix $\varrho_{\rm in}$) 
is subjected to a sequence of measurements, each of them drawn from the set 
of possible measurements $\{\MM_i\}$. The choice of the measurements is denoted 
by $x,y,z,...$ and the result by $a,b,c, ...$. This results in a probability 
distribution $p(abc ... | xyz ...)$. Note that if $x=y$ the measurement
$\MM_x$ is repeated two times, but this does not imply that the 
results $a$ and $b$ are the same.}
\label{fig:temporal1}
\end{figure}

What can we learn about quantum physics, if only a single quantum 
system is available? The only chance to obtain information  is to 
subject this quantum system (say, a single trapped ion) to a sequence 
of measurements and register the corresponding results. Here, measurements 
are procedures applied to the ion, resulting in a classical
outcome and a change of the ion's internal state. For a given set of 
 measurements there are different possible measurement sequences
of a certain length and these sequences may include repetitions of the
same measurement. Re-preparing the ion and repeating a sequence many times 
finally results in a probability distribution for a given sequence of 
measurements (see Fig.~\ref{fig:temporal1}).\\
\indent This probability distribution encodes the temporal correlations and 
these correlations can be used to violate Leggett-Garg inequalities 
\cite{leggett-garg-review, kofler} or to perform contextuality tests 
\cite{contextuality-tests}. Such tests can then prove that the quantum 
system violates certain assumptions of classicality and therefore they
have been intensively studied. For instance, one can ask for given 
correlations whether and at which cost they can be simulated classically 
\cite{kleinmann, brierley, fagundes, adansim}. This question may have important
implications in the characterization of quantum advantage in information processing tasks based on sequential
measurements \cite{Markiewicz}. 
Another question is what 
maximal correlations can be achieved within quantum mechanics 
\cite{fritz-temporal1, budroni, budroni-emary}. In these approaches, however, often assumptions 
must be made: For instance, contextuality tests require compatible measurements. For the special case of projective measurements, bounds on the maximal achievable temporal correlations have been provided \cite{budroni}. 
It remains unclear how to obtain such bounds for more general classes of measurements, as allowed by quantum mechanics where the post-measurement state may depend on 
the measurement result and the input state in a non-trivial way\cite{footnote}. This makes quantitative and analytical statements often difficult \cite{maurochsh}. 

In this paper we characterize temporal quantum correlations when measurements can be repeated, without making any assumption on the measurements. First, we study general correlations
without assuming the formalism of quantum mechanics, the only restriction 
being that later measurements do not influence the previous ones. 
The resulting probabilities form a polytope and we characterize 
its extreme points. Then, we relate this to the quantum mechanical formalism. 
We prove that any possible temporal correlation can be generated from measurements
on a quantum system, but the quantum system may be required to be high-dimensional. 
Interestingly, already for the simplest case of two measurements with 
two outcomes each, and measurement sequences of length two, there are temporal
correlations which cannot originate from quantum measurements on a qubit. This is 
unexpected, as the standard correlation for this case based on the CHSH 
inequality \cite{fritz-temporal1, budroni} does not have this 
property and all possible values for it can come from generalized measurements 
on a qubit \cite{budroni,maurochsh}. 
The correlations we found can then be used as tests of the quantum
dimension: We provide analytical bounds  without making any assumptions about the measurements and  a violation of these proves that 
the tested quantum system is three-dimensional. Finally, we discuss a possible
implementation using nitrogen-vacancy (NV) centers in diamond.

\section{The scenario} We consider temporal sequences of measurements 
on a quantum state $\varrho_{\rm in}$ (see Fig.~\ref{fig:temporal1}). 
In the simplest scenario, we consider two possible measurements with
two outcomes at two times. More precisely, at time $t_1$ one can 
choose between two different measurement settings $\MM_0$ and $\MM_1$. 
The input $x \in \{0,1\}$ determines which measurement is performed and 
the output is labelled by the variable $a \in \{0,1\}$. At a later time 
$t_2$, one can again choose between the two measurements $\MM_i$, where 
the input is labelled by $y$ and the output by $b$.  This leads to joint 
probabilities $p(ab \vert xy)$ for this measurement sequence. Note that for $x=y$ the same measurement is implemented at time $t_1$ and $t_2$, but different outcomes may be obtained. Clearly,
the probabilities obey the conditions of positivity, $p(ab\vert xy) \geq 0$ 
and normalization, $\sum_{a,b} p (ab\vert xy ) = 1.$ In addition, the first 
measurement is not influenced by the second one, so the probabilities fulfill 
the arrow of time (AoT) constraints \cite{kofler}
\begin{equation}\label{eq:cond1AoT}
\sum_b p(ab\vert xy) = \sum_b p(ab\vert xy') \quad \mbox{ for all } a,x, y, y'.
\end{equation}
This condition can be used to define the marginal probabilities $p(a\vert x)$
as
\begin{equation}
p\left(a\vert x\right) := \sum_b p\left(ab\vert xy\right) \quad \mbox{ for all } a,x,y.
\end{equation}
It is easy to see that the set of all probabilities forms a polytope. Furthermore, 
the definition and constraints can straightforwardly be generalized to 
sequences of arbitrary length $L$, number of results per measurement $R$
and number of possible measurement settings $S$ per time step. Again for any $R, S$ and $L$, the AoT constraints define a polytope, the so-called temporal correlation polytope, labelled by $P_L^{R,S}$. 

Before characterizing this polytope, let us review how the 
probabilities are determined in quantum mechanics. We consider the 
most general notion of measurements in quantum mechanics, which are 
described by quantum instruments (see, e.g., Ref.~\cite{teiko}). A measurement setting $\MM_s$ with the outcomes 
$\{r\}$ corresponds to a set of completely positive maps $\II_{r|s}$ which 
describe the state update and the probabilities. After measuring $\MM_s$ on the state $\varrho_{\rm in}$
and finding the result $r$ the (not normalized) post-measurement state is given by
\be
\varrho_{\rm out} = \II_{r|s}(\varrho_{\rm in}),
\ee
and the probabilities of obtaining the result is given by 
$p(r|s)= \tr[\II_{r|s}(\varrho_{\rm in})].$ Since some result must
occur, the maps $\II_{r|s}$ sum up to a trace-preserving map,
$\sum_r \II_{r|s} = \Lambda_s.$ If one is interested in the 
probabilities only, one can obtain them also by effects $\EE_{r|s}$. 
This means that for any measurement there are positive semidefinite 
operators $\EE_{r|s}$ which sum up to the identity 
$\sum_r \EE_{r|s} = \openone$ and which obey 
$p(r|s)= \tr(\EE_{r|s} \varrho_{\rm in})$. Finally, note that
in our formalism we do not consider a possible time evolution 
of the quantum system between the measurements. If there is such
a time evolution, it often can be absorbed in the positive maps $\II_{r|s}$,
as it just changes the post-measurement state.

\section{Characterizing the polytope}
Let us first mention a result on the structure of probability 
distributions $p(abc \dots \vert xyz \dots )$ that fulfill the 
AoT constraints.

\noindent
{\bf Observation~1.}
{\it A temporal probability distribution $p(abc \dots \vert xyz \dots)$ 
fulfills the AoT constraints if and only if it can be written as
\begin{equation}
\label{eqn-obs1}
p(abc \dots \vert xyz \dots) = 
p(a \vert x) p(b\vert axy) p(c\vert abxyz) \dots,
\end{equation}
with $p\left(a \vert x \right)$, $p(b \vert a x y)$, $p(c\vert abxyz)$ etc. being 
probability distributions with respect to the variables $a$, $b$, $c$ etc. 
}

Note that this is a straightforward generalization of an observation made in Ref.~\cite{fritz-temporal1}.
To understand this, note that for sequences 
of length two where $p(ab|xy)$ is given, one can define 
$p(b|axy)=p(ab|xy)/p(a|xy)=p(ab|xy)/p(a|x)$ and find the 
decomposition. On the other hand, if $p(b|axy)$ and $p(a|x)$ are 
given, Eq.~(\ref{eqn-obs1}) defines $p(ab|xy)$ and one can directly
verify all the desired properties. More details for longer sequences 
are given in Appendix A. 

The extreme points of $P_L^{R,S}$, i.e. the maximally achievable 
correlations can then be characterized as follows:

\noindent
{\bf Observation 2.}
{\it The extreme points of the temporal correlation polytope $P_L^{R,S}$ 
are given by the deterministic assignments that fulfill the AoT constraints.
Here, a deterministic assignment denotes a probability distribution that
takes either the value $0$ or $1$.
}

This Observation has been made independently in Ref.~\cite{jannik-thesis} 
and Ref.~\cite{abbott}. We present the proof from  Ref.~\cite{jannik-thesis} 
in Appendix A. Using the two observations, one can 
derive a simple equation to determine the number of extreme points 
$N_{L}^{R,S}$ for the polytopes $P_L^{R,S}$. We find
\begin{equation}
\label{eq:numb}
N^{R,S}_L =  \prod_{i=0}^{L-1} (R^S)^{(S^i)} 
= 
(R^S)^{(\frac{S^L -1}{S-1})}.
\end{equation}  
This formula can directly be understood for the simple case of 
$L=S = R=2$. To determine the number of extreme 
points of the polytope $P_2^{2,2}$, we first have to assign 
deterministic values for the marginal probabilities $p(a\vert x)$. 
For the probability distribution $p(a\vert x)$, there exist $R^S=2^2=4$ 
different deterministic assignments $p(a\vert x) = 0,1$. For the probability 
distribution of the second measurement we have, for a given first measurement,  
$R^S=2^2=4$ different deterministic assignments, but there are $S^1=2$ possible first 
measurement settings. This results in $2^2 \cdot (2^2)^2 = 64$ extreme 
points for the polytope $P_2^{2,2}$. In general, Eq.~(\ref{eq:numb}) can
be verified via induction and the geometric series, see also 
Appendix B. 


\section{Relation to quantum mechanics}
Having characterized the structure and number of the extreme points 
of $P_{L}^{R,S}$,  we will now clarify whether these extreme points 
can be reached by physical theories such as quantum mechanics. 

For simplicity, let us focus on the polytope $P_2^{2,2}$. Quantum mechanics 
fulfills the AoT constraints. In particular, the probabilities factorize via
\begin{align}
p(ab \vert xy) 
&= p(a \vert x) p(b\vert axy) 
\nonumber 
\\
&= \tr(\EE_{a\vert x} \vrho_{\rm in}) \tr(\EE_{b|y}\left[\frac{\II_{a|x}(\vrho_{\rm in})}{\tr(\EE_{a\vert x} \vrho_{\rm in})}\right]).
\end{align}
As stated in the following theorem all temporal probability distributions
can be reached with quantum mechanical measurements. Note that 
our result is different from a similar statement in Ref.~\cite{fritz-temporal1}, 
as we assume that measurements can be repeated and therefore some measurements 
must be represented by the same instrument (applied at different times) in 
the quantum mechanical description. 

\noindent
{\bf Theorem 1.}
{\it
Any probability distribution of $P_{L}^{R,S}$ can be reached in quantum mechanics.
The quantum mechanical representation may require a high-dimensional quantum
system.
}\\
\noindent{\it Proof.}  We consider the case of $P_{2}^{2,2}$, the construction can then be generalized to longer sequences.  
Let us first consider an extreme point. In order to reach this 
with measurements on a three-level system, i.e. a qutrit, we use the 
states $\ket{0}$, $\ket{1}$, and $\ket{2}$, where $\ket{0}$ is the 
initial state of the system, $\ket{1}$ is the post-measurement state 
after a first measurement with  $x=0$ and $\ket{2}$ is the 
post-measurement after the first measurement $x=1$.

For writing down suitable effects, we denote by $a_{0}^{x}$ the 
(deterministic) outcome when measuring $x$ at the first time step, 
by $a_{1}^y$ the outcome for the measurement $y$ at the second time 
step if one had measured $x=0$ and $a_{2}^y$ denotes the outcome of 
the second measurement if one had measured $x=1$. Note that 
$x,y$ are just labels for the first and second measurement, 
which have the same range of values. We have to define measurements
that can be carried out as a first {\it and} second measurement and 
denote the setting by $s$ and the result by $r$.

For each measurement $s\in \{0,1\}$ we define the set $\mathcal{P}_s$ which collects all 
the $i$ for which $a_i^s$ indicates a ``$0$'' result, i.e. $\mathcal{P}_s=\{i: a_{i}^s=0\}$. The effects of the measurements are then defined 
as $\EE_{0|s}= \sum_{i \in \mathcal{P}_s} \ketbra{i}$ and 
$\EE_{1|s}= \eins-\EE_{0|s}$. Note that with the initial and post-measurement states as defined before this ensures that the desired measurement outcomes are obtained. For writing down the complete instrument, 
we define $U_0=\ket{0}\bra{1}+\ket{1}\bra{0}+\ket{2}\bra{2}$
for the measurement $s=0$ and 
$U_1=\ket{0}\bra{2}+\ket{2}\bra{0}+\ket{1}\bra{1}$
for the measurement $s=1$, which will take care of the post-measurement states. That is we define the measurement 
operators $M_{r|s}=U_{s}\EE_{r|s}$ and the corresponding 
instrument $\mathcal{I}_{r|s} (\vr_{\rm in}) 
= M_{r|s} \vr_{\rm in} M_{r|s}^\dagger$. It can be easily seen that this reproduces the desired 
results if the system is initially prepared in $\ket{0}$. 
 
For the extension to non-extreme probability distributions in $P_{2}^{2,2}$
note first that in the previous construction the state 
and the effects depend on the extreme vertex, so one cannot 
directly obtain mixtures in  $P_{2}^{2,2}$ by mixing states and 
effects. But via increasing the dimension, it can be done as follows: 
Let ${E}$ be the set of all extreme points of the polytope 
$P_{2}^{2,2}$ and $\{\ket{0_e}, \ket{1_e},\ket{2_e}\}_{e\in {E}}$ 
be the orthonormal vectors for the extreme vertex as constructed above.
We also denote the measurement operators as $M_{r|s}^e$, stressing the
dependence on the vertex $e \in E$. If we have a general set of probability distributions 
$\mathfrak{p}$ in $P_{2}^{2,2}$ we can write it as $\mathfrak{p} = 
\sum_e \alpha_e e$ with probabilities $\alpha_e$ and $ e \in {E}$.
Then, this can be reproduced by taking as an initial state the direct 
sum
$\vr=\bigoplus_{e\in {E}} \alpha_e \ketbra{0_e}$
and as measurement operators 
$M_{r|s}=\bigoplus_{e\in E} M_{r|s}^e.$ 
$\hfill \Box$

Note that the protocol that allows us to obtain any probability distribution in $P_{2}^{2,2}$ corresponds to a completely classical strategy as it does not require any coherences. It can be shown that for $R=S=2$ there exist extreme points which require a minimal dimension to be reached that scales at least as $2^{L+1}/L$ \cite{generaltempsequ}.


\section{Temporal correlations for a qubit} The previous construction allows us to reach all extreme points of the polytope $P^{2,2}_2$ using a three-level quantum system. However, already in the simple scenario
of $P^{2,2}_2$ not all extreme points can be reached using a qubit
only. Among the 64 extreme vertices many are equivalent, as one can
relabel the measurement settings $0 \leftrightarrow 1$ and the
measurement outcomes $0 \leftrightarrow 1$ \cite{footnote2}. Taking these symmetries into
account, 10 extreme vertices remain, and 6 of these can be reached via
measurements on a qubit in a simple manner \cite{jannik-thesis}. The 
four remaining extreme points are
\begin{align}
e_1: \; p(00|00) = p(00|11) = p(01|01) = p(01|10) = 1,
\nonumber
\\
e_2: \; p(01|00) = p(01|11) = p(00|01) = p(00|10) = 1,
\nonumber
\end{align}
\begin{align}
e_3: \; p(01|00) = p(00|11) = p(01|01) = p(01|10) = 1,
\nonumber
\\
e_4: \; p(01|00) = p(01|11) = p(01|01) = p(00|10) = 1.
\end{align}

It is instructive to discuss these vertices in a qualitative manner, 
leading to an intuitive understanding why they cannot be
reached by measurements on a qubit.  Let us first discuss the 
vertex $e_1$. From $p(01|01) = p(01|10) = 1$ it follows that 
the measurements  $\MM_0$ and $\MM_1$ cannot be trivial in the 
sense that their result must depend on the input state. Then, the 
probabilities $p(00|00) = p(00|11)=1$ can only be reached on a qubit 
if the effects of the measurements $\MM_0$ and $\MM_1$ are of the form $\EE_{0|i}=\ketbra{a_0}+p_i \ketbra{a_1}$ and $\EE_{1|i}=(1-p_i) \ketbra{a_1}$ with $0\leq p_i<1$ and $\{\ket{a_0},\ket{a_1}\}$ being an orthonormal basis and the input state is $\ket{a_0}$. Moreover, 
it follows that the measurements do not disturb this input state. But then, a contradiction
to $p(01|01) = p(01|10) = 1$ occurs. 

Concerning the vertex $e_3$, as already mentioned the fact that $p(01|01) = p(01|10) = 1$ implies that the outcome of $\mathcal{M}_1$ (and $\mathcal{M}_0$) has to depend on the state on which the measurement is performed. It follows then from $p(00|11) =1$ that $\mathcal{M}_1$ has effects of the form $\EE_{0|1}=\ketbra{a_0}+p_1 \ketbra{a_1}$ and $\EE_{1|1}=(1-p_1) \ketbra{a_1}$ with $0\leq p_1<1$ and the input state is $\ket{a_0}$ which remains unchanged by measuring $\mathcal{M}_1$. However, considering $p(01|00) =p(01|10) = 1$ this leads to a contradiction.
The discussion of the vertices $e_2$ and $e_4$ is similar, 
further details and rigorous proofs can be found in Appendix~D. 

\section{Dimension witnesses}
The fact that the vertices can not be represented by qubit systems 
can be made quantitative by characterizing the amount to which they 
can be approximated. For that we consider two expressions, derived 
from the vertices $e_1$ and $e_3$, 
\begin{align}
\BB_{1}= p(00|00) +p(00|11) + p(01|01) + p(01|10),
\nonumber
\\
\BB_{3}=p(01|00) + p(00|11) +p(01|01) + p(01|10).
\end{align}
For these expectation values we can state: 

\noindent
{\bf  Theorem 2. } {\it For arbitrary measurements on a qubit the
following bounds hold:
\begin{align}
\BB_{1} & \leq \mathcal{C}_1 = 3,
\nonumber
\\
\BB_3 & \leq \mathcal{C}_3 \approx 3.186.
\end{align}
The bound $\mathcal{C}_3$ is given by the root of a polynomial of degree ten.}

\noindent{\it Proof.} The proof is given in Appendix C.

The first question that arises is whether similar bounds can be established for the vertices $e_2$ and $e_4$. We first note that the structure of the vertices $e_1$ and $e_2$ is similar and that the same holds true for the vertices $e_3$ and $e_4$. For the vertices $e_2$ and $e_4$ one can analytically reduce the problem of finding the maximal expectation
value for the corresponding $\BB_2$ and $\BB_4$, to a maximization that depends only on two parameters. Performing the remaining maximization numerically suggests that $\BB_2$ is upper bounded by $ \mathcal{C}_1$ and $\BB_4$ is upper bounded by $\mathcal{C}_3$. This reflects the similarity among the pairs of vertices mentioned above. For the vertex $e_4$ one can further analytically show that $\BB_4$ is upper bounded by $2+\sqrt{2}$.

It is instructive to characterize the qubit measurements giving
the maximal value for $\BB_1$ and $\BB_3.$ For $\BB_1$ the optimal measurement
$\MM_0$ is a trivial measurement, giving always the result $0$. Then, if 
$\MM_1$ is a $\sigma_z$ measurement, the initial state is $\ket{0}$ and
$\MM_0$, although giving a fixed result, flips the state, a value of 
$\BB_1 =3$ can be reached. For $\BB_3$ the optimal measurements can be shown to have projective effects. It should be noted that for such measurements $\BB_3$ is equivalent to $\BB_4$.  Note further that from the proof of Theorem 1 it follows 
that $\BB_1 = \BB_3 = 4$ can be reached on a three-level system.

Finally, the preceding Theorem shows that temporal correlations can 
be used for characterizing the dimension of quantum systems in a 
device-independent manner. Namely, if one of the inequalities in 
Theorem 2 is violated, one knows that the underlying system is not 
a qubit, without assuming anything about the measurements or the state transformations between the measurements \cite{footnote3}. So far, 
dimension tests have been proposed using Bell inequalities \cite{brunner}, 
prepare \& measure schemes \cite{gallego} or more general input-output correlations \cite{dallarno1, dallarno2}, the time evolution of the expectation value of a single observable \cite{wolf} or temporal correlations and  
contextuality \cite{guehnedim, fritz-temporal1, budroni-emary}. The schemes using Bell 
inequalities cannot be used with a single qutrit and moreover, recently 
it turned out that their violation can also be observed with pairs of 
qubit systems \cite{cong, tristan}. The approach in Ref.~\cite{dallarno1}, based on a result on noiseless $n$-level quantum channels \cite{frenkel}, is, however, restricted to some specific channels, which are inserted between the preparation and the measurement. In Ref.~\cite{dallarno2}, the results are not limited to specific measurement, but they are restricted to the qubit case. Such approaches have been further developed to propose a 
principle, based on temporal correlations, that may single out quantum theory among generalized probabilistic theories \cite{dallarno3}, and to provide a characterization of quantum memories based on temporal correlations \cite{rosset}. The existing proposals using temporal 
correlations \cite{guehnedim, fritz-temporal1, budroni-emary} make assumptions about 
the nature of the measurements, e.g., they have to be projective. The dimension bound derived in \cite{wolf} is based on the expectation values of a single observable evaluated after $t$ uses of a quantum channel and assumes that the time evolution is Markovian and homogeneous in time.

The dimension witness from Theorem~2 is closest to the prepare \& measure 
scheme from Ref.~\cite{gallego}, which is also device-independent. As any 
measurement can be viewed as a state preparation, our scheme may also be 
interpreted as prepare \& measure scheme, but there are several advantages
of our approach: First, the bound for $\BB_1$ leads to larger separation
among qutrits and qubits than some of the inequalities in Ref.~\cite{gallego}.
Second, our approach can be generalized to measurement sequences of length 
three or longer, and then an interpretation as a prepare \& measure scheme 
is not possible anymore.

Our scheme does not only provide a distinction between a qubit and higher-dimensional systems but also allows us to establish a lower bound on the distance between the measured system and a qubit. In order to quantify this distance one may define $(d+\epsilon)$-dimensional systems as systems for which the initial states as well as all possible post-measurement states of the instruments deviate only by $\epsilon$ from the same d-dimensional subspace. More precisely, a $(d+\epsilon)$- system is given by an initial state $\vrho_{\rm in}$ and instruments $\{\II_{a|x}\}_a$ with the property that there exists a projector on a d-dimensional subspace, $P_d$, such that $\lVert P_d\vrho_{\rm in}P_d-\vrho_{\rm in} \rVert_{\rm tr}\leq \epsilon$ and for all $a, x$ and quantum states $\rho$ it holds that $\lVert P_d\II_{a|x}(\rho)P_d-\II_{a|x}(\rho) \rVert_{\rm tr}\leq \epsilon$. It can be shown that for $(2+\epsilon)$-dimensional systems it holds that $\mathcal{B}_i \leq C_i+12 \epsilon$ (see Appendix D). 
Hence, from the value of $\mathcal{B}_i$ determined in an experiment it is possible to deduce a lower bound on $\epsilon$.

\section{Possible implementation with NV centers} 
NV centers are well-characterized quantum systems \cite{nv-general} and, 
since they contain several energy levels, they are candidates for testing 
the inequalities derived above. 

The relevant energy levels of an NV cente are a ground state manifold 
$^3$A and a set of excited states $^3$E. Both manifolds consist of three 
states, corresponding to the $m_s=0$ and $m_s=\pm 1$ quantum numbers. The 
ground state manifold can be used as a qutrit: In the presence of a 
magnetic field they are non-degenerate and at low temperatures microwave 
fields can be used to drive unitary transitions between the three states. 
Coupling the ground state manifold $^3$A to the manifold $^3$E with resonant 
excitation preserves the value of $m_s$, but as the $m_s=\pm 1$ state in $^3$E 
decays with some probability to the $m_s=0$ state in $^3$A, the ground state 
manifold can be prepared in the state $m_s=0$ with high fidelity \cite{robledo}. 
By driving the transition from $m_s=0$ in $^3$A to $m_s=0$ in $^3$E only,
and detecting fluorescence, the $m_s=0$ state can be read out and at low
temperatures the $m_s=\pm 1$ states remain unaffected \cite{robledo}. 

Our procedure of reaching the extreme vertex $e_1$ using a three-level
system (see the proof of Theorem 1) leads to the following operators: Initially, 
the NV center is prepared in the $\ket{0} = \ket{m_s=0}$ state. The measurement
$\MM_0$ is performed in the following way: First, the NV center is measured 
projectively in the $\ket{2}=\ket{m_s=-1}$ state (result ``1'') or in the 
orthogonal subspace, spanned by $\ket{0}$ and $\ket{1}=\ket{m_s=1}$ (result ``0''). 
This first step can be achieved by first applying a unitary transformation, 
then performing the projective measurement of $\ket{0}$ and finally undoing 
the unitary again. The second step of the measurement $\MM_0$ is a unitary 
transformation $U_0=\ket{0}\bra{1}+\ket{1}\bra{0}+\ket{2}\bra{2}$ independent 
of the measurement result. The measurement $\MM_1$ is implemented similarly, 
first one projects onto $\ket{1}$ (result ``1'') or the orthogonal subspace 
(result ``0'') and finally one performs 
$U_1=\ket{0}\bra{2}+\ket{2}\bra{0}+\ket{1}\bra{1}.$ 
These measurements will lead to $\BB_1=4$ which is the maximal violation 
of the inequality $\BB_1\leq 3.$


\section{Conclusion}
We have characterized general temporal correlations coming from sequences of 
measurements on a quantum system. We first considered general correlations
obeying the arrow-of-time condition and showed that all of these can be attained
by quantum mechanics. If the dimension of the quantum system is restricted, however, 
not all correlations can arise from quantum mechanical systems. This allows us to 
construct dimension witnesses, which can be implemented with NV centers in diamond. 

There are several directions in which our approach can be generalized. First, it 
would be interesting to characterize the set of quantum correlations coming from 
a fixed dimension further. To give an example of an open question, it is not 
clear whether this set it convex. Second, it is promising to consider longer 
measurement sequences for dimension tests. This will probably lead to higher 
violations and easier experimental implementation. Finally, it is important to 
understand the classical protocols to generate temporal correlations. For 
instance, classical systems with a bounded memory can also not reproduce
all temporal correlations, and this may help to characterize the quantum advantage
in information processing based on sequential measurements.\\ \\
We thank Johannes Greiner, Matthias Kleinmann and Gael Sent\'{i}s  for 
discussions. This work was supported by the DFG, the Austrian Science Fund (FWF):
M 2107 (Meitner-Programm) and the ERC (Consolidator 
Grant No. 683107/TempoQ).

\section{Appendix A: Proof of the Observations 1 and 2}
In this part of the Appendix, we will present the proofs of Observations 
1 and 2. We start with the proof of Observation 1 which we will use 
afterwards to prove Observation 2.

\noindent
{\bf Observation~1.}
{\it A temporal probability distribution $p(abc \dots \vert xyz \dots)$ 
fulfills the AoT constraints if and only if it can be written as
\begin{equation}
\label{eqn-obs1app}
p(abc \dots \vert xyz \dots) = 
p(a \vert x) p(b\vert axy) p(c\vert abxyz) \dots,
\end{equation}
with $p\left(a \vert x \right)$, $p(b \vert a x y)$, $p(c\vert abxyz)$ etc. being 
probability distributions with respect to the variables $a$, $b$, $c$ etc. 
}

{\it Proof.} 
We will prove this Observation for the case $L=3$, as it is straightforward to 
generalize the proof to sequences of arbitrary length. First, assume that 
the probability distribution $p(abc\vert xyz)$ fulfills the AoT constraints. 
In this case, the marginal probabilities $p(a\vert x)$ are well defined. 
We will further introduce two conditional probabilities $p(b\vert axy)$ 
and $p(c\vert abxyz)$. We define $p(b\vert axy)$ as
\begin{equation}
p\left(b\vert axy\right) := \dfrac{p\left(ab\vert xy\right)}{\sum_b p\left(ab\vert xy\right)} = \dfrac{p\left(ab\vert xy\right)}{ p\left(a\vert x\right)},
\end{equation}
for $p\left(a\vert x\right) \neq 0$, and one can, for example, choose
\begin{equation}
p\left(b\vert axy\right):=0,
\end{equation}
for $p\left(a\vert x\right) = 0$.
It is easy to see that $p(b\vert axy)$ is a valid probability distribution, 
if $p(a\vert x)\neq 0$. First $p(b\vert axy)$ is positive since $p(ab\vert xy)$ 
and $p(a\vert x)$ are positive and second we have
\begin{equation}
\sum_b p(b\vert axy) = \dfrac{\sum_b p(ab\vert xy)}{p(a\vert x)} = 1.
\end{equation}
The conditional probability $p(c\vert abxyz)$ is defined as
\begin{equation}
p(c\vert abxyz) = \dfrac{p(abc\vert xyz)}{p(ab\vert xy)},
\end{equation}
if $p(ab\vert xy) \neq 0$ and can be chosen to be zero otherwise. 
In an analogous way as for $p(b\vert axy)$, we can show that 
$p(c\vert abxyz)$ is a valid probability distribution if 
$p(ab\vert xy) \neq 0$. 
We then simply have 
\begin{align}
p\left(abc\vert xyz\right)&= p\left(a\vert x\right) 
\dfrac{p\left(ab\vert xy\right)}{p\left(a\vert x\right)} 
\dfrac{p\left(abc\vert xyz\right)}{p\left(ab\vert xy\right)}
\nonumber
\\
 &= p\left(a\vert x\right) p\left(b\vert axy\right) p\left(c\vert abxyz\right).
\end{align}
Hence the probability distribution $p(abc\vert xyz)$ can be factorized if 
it fulfills the AoT constraints.

Now assume that we have probability distributions $p(a\vert x)$, $p(b\vert axy)$ 
and $p(c\vert abxyz)$. The product of these probabilities
\begin{align}
p(a\vert x)p(b\vert axy)p(c\vert abxyz)=:p(abc\vert xyz),
\end{align}
fulfills the AoT constraints, as
\begin{align}
 \sum_{b,c} p\left(abc\vert xyz\right) &= \sum_{b,c}p\left(a\vert x\right)p\left(b\vert axy\right) p\left(c\vert abxyz\right)
 \nonumber
 \\
&= p\left(a\vert x\right)\sum_b p\left(b\vert axy\right)
\sum_c p\left(c\vert abxyz\right)
\nonumber
\\
&= p\left(a\vert x\right) \quad \forall a,x,y,z,
\end{align}
and 
\begin{align}
 \sum_{c} p\left(abc\vert xyz\right) 
 &= \sum_{c}p\left(a\vert x\right) p\left(b\vert axy\right) p\left(c\vert abxyz\right) 
 \nonumber
 \\
&= p\left(a\vert x\right) p\left(b\vert axy\right)\sum_c p(c\vert abxyz)
\nonumber
\\
&= p\left(a\vert x\right)  p\left(b\vert axy\right) 
\nonumber
\\
&=: p\left(ab\vert xy\right)\quad\forall a,b,x,y,z.
\end{align}
So the probability distribution 
$p\left(abc\vert xyz\right) = p\left(a\vert x\right) p\left(b\vert axy\right) p\left(c\vert abxyz\right)$ fulfills the AoT constraints, which completes the proof.
$\hfill \Box$

Using Observation 1, we can now prove Observation 2:

\noindent
{\bf Observation 2.}
{\it The extreme points of the temporal correlation polytope $P_L^{R,S}$ 
are given by the deterministic assignments that fulfill the AoT constraints.
Here, a deterministic assignment denotes a probability distribution that
takes either the value $0$ or $1$.}

{\it Proof.} 
We need to show that 
\begin{itemize}
\item[(i)] All deterministic assignments are extreme
\item[(ii)] Every vector $v$ consisting of the probabilities $p(abc...\vert xyz...)$ 
can be written as a convex combination of the vectors corresponding to deterministic 
assignments.
\end{itemize}
The proof of (i) is trivial in the sense that a deterministic assignment for the 
vector $v$ can never be written as a convex combination of other vectors. We will 
show (ii) for the polytope $P_2^{2,2}$, however, one can easily generalize the method 
to an arbitrary polytope $P_L^{R,S}$.

Let us first consider the probabilities $p(a\vert x)$ 
for fixed $x$. Let us define the vector 
\begin{align}\label{eq:convex1}
v_x &= \left(\begin{array}{c}
p(0\vert x) \\ 
p(1\vert x)
\end{array}\right) = c \left(\begin{array}{c}
1 \\ 
0
\end{array}\right) + 
(1-c) \left(\begin{array}{c}
0 \\ 
1
\end{array}\right) 
= \left(\begin{array}{c}
c\\ 
1-c
\end{array}\right), 
\end{align}
which is a convex combination of  $(1,0)^T$ and 
$(0,1)^T$, describing probability $1$ for outcome $0$ and probability 
$1$ for outcome $1$, respectively.

Let us assume first that $c\neq 0,1$. Then, for fixed $x$ and 
$y$, the vector containing the conditional probabilities $p(b\vert axy)$ 
can also be written as a convex combination. In this case, we define 
the vector
\begin{align}\label{eq:convex2}\notag
v_{xy} &= \left(\begin{array}{c}
p(0\vert 0xy) \\ 
p(1\vert 0xy) \\ 
p(0\vert 1xy) \\ 
p(1\vert 1xy)
\end{array} \right) = d\left(\begin{array}{c}
1 \\ 
0 \\ 
0 \\ 
0
\end{array} \right) + (1-d)\left( \begin{array}{c}
0 \\ 
1 \\ 
0 \\ 
0
\end{array}\right) \\
&+ e\left(\begin{array}{c}
0 \\ 
0 \\ 
1 \\ 
0
\end{array}\right) + (1-e)\left( \begin{array}{c}
0 \\ 
0 \\ 
0 \\ 
1
\end{array} \right) = \left(\begin{array}{c}
d \\ 
1-d \\ 
e \\ 
1-e
\end{array} \right).  
\end{align} 
The first two entries describe the probability distribution for $a=0$, with the convex coefficient $d$ and the last two, the probability distribution for $a=1$ with the convex coefficient $e$. Both convex combinations are independent of each other. 

We now want to show that for fixed $x$ and $y$ the probability distribution $p(ab\vert xy)$ is always a convex combination if the conditional probabilities are non-deterministic. For this, let us define the vector
\begin{equation}
v = \left(\begin{array}{c}
p(00\vert xy) \\ 
p(01\vert xy) \\ 
p(10\vert xy \\ 
p(11\vert xy)
\end{array} \right) = \left(\begin{array}{c}
p(0\vert x)p(0\vert 0xy) \\ 
p(0\vert x) p(1\vert 0xy) \\ 
p(1\vert x) p(0\vert 1xy \\ 
p(1\vert x)p(1\vert 1xy)
\end{array} \right),
\end{equation}
where we used the fact that due to Observation 1, we can factorize the probabilities 
$p(ab\vert xy)$ into the conditional probabilities $p(a\vert x)$ and $p(b\vert axy)$. 
If we replace the probabilities $p(a\vert x)$ and $p(b\vert axy)$ with the respective 
coefficients in the vectors in Eqs.~(\ref{eq:convex1}) and (\ref{eq:convex2}), we obtain
\begin{align}\small
v &= \left(\begin{array}{c}
cd \\ 
c(1-d) \\ 
(1-c)e \\ 
(1-c)(1-e)
\end{array} \right) 
= cd\left(\begin{array}{c}
1 \\ 
0 \\ 
0 \\ 
0
\end{array} \right) + c(1-d) \left(\begin{array}{c}
0\\ 
1 \\ 
0 \\ 
0
\end{array} \right) 
\nonumber
\\
&+ (1-c)e\left(\begin{array}{c}
0 \\ 
0 \\ 
1 \\ 
0
\end{array} \right) + (1-c)(1-e) \left(\begin{array}{c}
0 \\ 
0 \\ 
0 \\ 
1
\end{array} \right),
\end{align}
which is a convex combination of at least two different vectors if 
$0< c < 1$, $0\leq d\leq 1$ and $0\leq e\leq 1$, since 
$cd, c(1-d),(1-c)e, (1-c)(1-e) \geq 0$ and $cd+c(1-d)+(1-c)e+(1-c)(1-e)=1$.

Up to now, we restricted ourselves to the case, where $p(a\vert x) \neq 0$ 
for all $a$, given $x$.
Consider next the case of a deterministic assignment for $p(a\vert x)$ and 
fixed $x$, i.e. $c=0$ or $c=1$. Assume without loss of generality that $c=1$. The vector $v$ is then of the form
\begin{equation}
v = \left(\begin{array}{c}
d \\ 
(1-d) \\ 
0 \\ 
0
\end{array} \right) = d\left(\begin{array}{c}
1 \\ 
0 \\ 
0 \\ 
0
\end{array} \right) + (1-d) \left(\begin{array}{c}
0\\ 
1 \\ 
0 \\ 
0
\end{array} \right),
\end{equation}
which is still a convex combination of deterministic assignments.

Since we can construct vectors like this for every choice of $x$ and $y$, 
we find that all non-deterministic assignments for the vector $v$ are convex 
combinations of deterministic assignments. $\hfill \Box$

\section{Appendix B: On the number of extreme vertices}
In this part of the Appendix, we will present the proof of Eq.~(\ref{eq:numb}), which 
quantifies the number of extreme points of $P_L^{R,S}$.

We prove this equation by induction. For sequences of length $L=1$, we have $R^S$ 
different possibilities to assign a deterministic value to the probabilities 
$p(a\vert x)$, i.e $N_1^{R,S} = \prod_{i=0}^{0} (R^S)^{(S^i)}= R^S.$

Next we show that the equation is true for a sequence of length $L+1 $, under the assumption 
that it is valid for sequences of length $L$. Given a specific sequence of measurements of 
length $L$ there are  $R^S$ different deterministic assignments possible for the probability 
distribution of the measurements at time $L+1$. Moreover, the number of possible measurement 
settings for a sequence of length $L$ is given by $S^L$. This leads to  $(R^S)^{S^l}$ different 
possible deterministic assignments for the probability distributions of the measurements at 
time $L+1$ of an extreme point. Hence, the total number of different deterministic assignments 
for sequences of length $L+1$ is given by $N_L^{R,S}\cdot (R^S)^{S^L}=N_{L+1}^{R,S}$. 
With this we have shown that
\begin{equation}
N^{R,S}_L =  \prod_{i=0}^{L-1} (R^S)^{(S^i)} 
= (R^S)^{(\frac{S^L -1}{S-1})},
\end{equation}  
where for the second equality the well known formula for the partial sum of 
the geometric series has been used.
$\hfill \Box$

\section{Appendix C: Proof of Theorem 2}

This part of the appendix is concerned with the proof of Theorem 2. As mentioned in the main text, there are four (up to relabeling of the measurement outcomes and settings) extreme points of $P_2^{2,2}$ that cannot be reached via measurements on a qubit:
\begin{align}
e_1: \; p(00|00) = p(00|11) = p(01|01) = p(01|10) = 1,
\nonumber
\\
e_2: \; p(01|00) = p(01|11) = p(00|01) = p(00|10) = 1,
\nonumber
\\
e_3: \; p(01|00) = p(00|11) = p(01|01) = p(01|10) = 1,
\nonumber
\\
e_4: \; p(01|00) = p(01|11) = p(01|01) = p(00|10) = 1.
\end{align}
In order to quantify to which amount they can be approximated we introduced in the main text the quantities
\begin{align}
\BB_{1}= p(00|00) +p(00|11) + p(01|01) + p(01|10),
\nonumber
\\
\BB_{3}=p(01|00) + p(00|11) + p(01|01) + p(01|10),
\end{align}
for which one can can find upper bounds for two-dimensional systems as stated in Theorem 2:\\

\noindent
{\bf  Theorem 2. } {\it For arbitrary measurements on a qubit the
following bounds hold:
\begin{align}
\BB_{1} & \leq \mathcal{C}_1 = 3,
\nonumber
\\
\BB_3 & \leq \mathcal{C}_3 \approx 3.186.
\end{align}The bound $\mathcal{C}_3$ is given by the root of a polynomial of degree ten.}

In the following subsections we will discuss the inequalities associated to the extreme points in more detail and provide a proof of Theorem 2.
\subsubsection{The extreme point $e_1$ and its associated temporal inequality}
In this subsection we show that for arbitrary measurements on a single qubit the quantity $\BB_{1}= p(00|00) +p(00|11) + p(01|01) + p(01|10)$ is smaller or equal to $3$. Moreover, we show that this bound is tight, i.e. there exists a sequence of measurements on a qubit that allows  to reach this value.

\noindent{\it Proof.} We will first show that for all initial states and post-measurement states the maximal value of $\BB_{1}$ is either smaller or equal to 3 [case (a)] or is attained in case the effects for both measurements are projectors [case (b)]. We will then consider case (b). We will  identify the optimal initial and post-measurement states for such measurements and show that the maximal value for $\BB_{1}$  that can be obtained with projective effects is given by $3/2+\sqrt{2}<3$. We finally show that the upper bound given by $3$ is tight by providing an explicit protocol that allows to reach  $\BB_{1}=3$. 

Let us start by defining our notation. Throughout this proof the effects for $\mathcal{M}_0$ ($\mathcal{M}_1$) corresponding to the outcome $r\in \{0,1\}$ will be denoted by $\EE_{r|0}$ ($\EE_{r|1}$) respectively. We will first use the the following decomposition for these effects,
\begin{align}\label{effapp210}
&\EE_{0|0}=a_0(\openone + b_0\,\vec{c}\cdot\vec{\sigma}),\\\label{effapp220}
&\EE_{1|0}=\openone-\EE_{0|0},\\
&\EE_{0|1}=a_1(\openone + b_1\,\vec{d}\cdot\vec{\sigma}),\\\label{effapp240}
&\EE_{1|1}=\openone-\EE_{0|1},
\end{align}
where $\vec{c},\vec{d}\in\mathbb{R}^3$, $|\vec{c}|=|\vec{d}|=1$ and $\vec{\sigma}=(\sigma_1,\sigma_2,\sigma_3)$ with $\sigma_i$ being the Pauli matrices. Moreover, we choose without loss of generality that $b_s\geq 0$ and therefore due to $\openone\geq \EE_{r|s}\geq 0$ we have that $0\leq a_s\leq \frac{1}{1+b_s}$ and $b_s\leq 1$ for $s\in\{0,1\}$.
Note that due to the AoT constraints we have that $p(ab|xy)=p(a|x)p(b|axy)$ and therefore
\begin{eqnarray}\nonumber
\BB_{1}&=p(0|0)[p(0|000)+p(1|001)]\\&+p(0|1)[p(1|010)+p(0|011)].
\end{eqnarray}
In the following we will denote by $\vrho_{\rm in}$ the initial state and by $\vrho_{x}$ the post-measurement states given that measurement $\mathcal{M}_x$ has been performed at time $t_1$ and outcome $0$ has been obtained.  We will use for these states their Bloch decomposition \bea\label{eqrho0}
\vrho_j=\frac{1}{2}(\openone+\vec{\alpha}_{j}\cdot\vec{\sigma})
\eea
for $j\in\{{\rm in}, 0, 1\}$ where $\vec{\alpha}_{j}\in\mathbb{R}^3$. Note due to the fact that $\vrho_j$ has to be positive semidefinite we have that $|\vec{\alpha}_{j}|\leq 1$. 
Using the decomposition for the effects in Eqs.(\ref{effapp210})-(\ref{effapp240})  one obtains that 
\begin{align}
&p(0|0)=a_0\,(1+b_0\vec{c}\cdot\vec{\alpha}_{\rm in}),\\
&p(0|1)=a_1\,(1+b_1\vec{d}\cdot\vec{\alpha}_{\rm in}),\\\
&p(0|000)=a_0\,(1+b_0\vec{c}\cdot\vec{\alpha}_{0}),\\
&p(1|001)=1-a_1\,(1+b_1\vec{d}\cdot\vec{\alpha}_{0}),\\
&p(1|010)=1-a_0\,(1+b_0\vec{c}\cdot\vec{\alpha}_{1}),\\
&p(0|011)=a_1\,(1+b_1\vec{d}\cdot\vec{\alpha}_{1}).
\end{align}
We will first show that for any $\vrho_{\rm in}$, $\vrho_{0}$, $\vrho_{1}$ the maximum of $\BB_{1}$ is smaller or equal to 3 or is attained for $a_0=\frac{1}{1+b_0}$. In order to do so we first consider $\BB_{1}$ as a function of $a_0$ (all other parameters are assumed to be fixed but arbitrary) and derive its critical points. The derivative of $\BB_{1}$ with respect to $a_0$ is given by
\begin{align}\nonumber
\frac{d\BB_{1}}{d a_0}&=[p(0|000)+p(1|001)]\,(1+b_0\vec{c}\cdot\vec{\alpha}_{\rm in}) \\ &+p(0|0)\,(1+b_0\vec{c}\cdot\vec{\alpha}_{0})-p(0|1)\,(1+b_0\vec{c}\cdot\vec{\alpha}_{1}).
\end{align}
Multiplying this equation by $a_0$, one obtains at the critical points that 
\begin{align}\nonumber
&[p(0|000)+p(1|001)]p(0|0)\\ &= p(0|1)[1-p(1|010)]-p(0|0)p(0|000)\leq 1.
\end{align}
Note that this implies that  $\BB_{1}$ at the points where the derivative vanishes cannot exceed 3. We also have to consider the boundary of the domain for $a_0$, i.e. we have to consider $a_0=0$ and $a_0=\frac{1}{1+b_0}$.
As can be easily seen  $\BB_{1}\leq 2$ for $a_0=0$. In order to investigate the case $a_0=\frac{1}{1+b_0}$ in more detail we will use that in this case the effects of the measurements in Eqs.(\ref{effapp210}) and (\ref{effapp220})  can equivalently (by substituting $b_0=\frac{p}{2-p}$) be written as 
\begin{align}\label{effapp10}
&\EE_{0|0}=\frac{1}{2}[(2-p) \openone + p \,\vec{c}\cdot\vec{\sigma}],\\\label{effapp40}
&\EE_{1|0}=\frac{p}{2}(\openone -\, \vec{c}\cdot\vec{\sigma}),
\end{align} where $0\leq p\leq 1$. 
Considering now  $\BB_{1}$ as a function of $p$ (and again assuming all other parameters as fixed but arbitrary) one obtains for its derivative
\begin{align}\nonumber
\frac{d\BB_{1}}{d p}&=-[p(0|000)+p(1|001)]\,\frac{1}{2}(1-\vec{c}\cdot\vec{\alpha}_{\rm in}) \\ &-p(0|0)\,\frac{1}{2}(1-\vec{c}\cdot\vec{\alpha}_{0})+p(0|1)\,\frac{1}{2}(1-\vec{c}\cdot\vec{\alpha}_{1}).
\end{align}
Hence, at the critical points we have that 
\begin{align}\nonumber
0&=[p(0|000)+p(1|001)][p(0|0)-1]\\\nonumber &+p(0|0)[p(0|000)-1]+p(0|1)p(1|010)\\\nonumber &\geq 2 [p(0|0)-1]+p(0|0)[p(0|000)-1]\\ &+p(0|1)p(1|010),
\end{align}
where we used that $p(0|000)+p(1|001)\leq 2$ and $p(0|0)-1\leq 0$.
This implies that 
\begin{align}\nonumber 2&\geq p(0|0)[p(0|000)+1]+p(0|1)p(1|010)\\
&\geq p(0|0)[p(0|000)+p(1|001)]+p(0|1)p(1|010)
\end{align}
and therefore it holds that $\BB_{1}\leq 3$ at the points where this derivative vanishes. We will next consider the boundary of the domain of $0\leq p\leq 1$. It is straightforward to see that for $p=0$ one obtains that $\BB_{1}\leq 3$. Before we investigate the case $p=1$ in more detail we will use that $\BB_{1}$ is invariant under exchanging measurement $0$ and measurement $1$. Therefore, one can analogously show that if the effects of measurement $1$ are not projectors we have that $\BB_{1}\leq 3$ and it only remains to show that in the case of projective effects for both measurements it also holds true that $\BB_{1}\leq 3$.
In the following we we will use the notation
\begin{align}
&\EE_{0|1}=\frac{1}{2}(\openone + \,\vec{d}\cdot\vec{\sigma}),\\
&\EE_{1|1}=\frac{1}{2}(\openone -\vec{d}\cdot\vec{\sigma}).
\end{align} 
With this, Eq. (\ref{eqrho0}) and Eqs. (\ref{effapp10})-(\ref{effapp40}) for $p=1$ we have that $\BB_{1}$ is of the following form
\begin{align}\nonumber\BB_{1}&=\frac{1}{4}(1+\, \vec{c}\cdot\vec{\alpha}_{\rm in})[2+(\vec{c}-\vec{d})\cdot\vec{\alpha}_{0}]\\\label{eqf0}
&+\frac{1}{4}(1+\vec{d}\cdot\vec{\alpha}_{\rm in})[2+(\vec{d}-\vec{c})\cdot\vec{\alpha}_{1}].
\end{align}
As $\frac{1}{2}(1+\, \vec{c}\cdot\vec{\alpha}_{\rm in})=p(0|0)\geq 0$ the optimal choice of  $\vec{\alpha}_{0}$ is given by
\bea
\vec{\alpha}_{0}= \frac{\vec{c}-\vec{d}}{\sqrt{2-2\cos (\gamma )}}\,\, \mathrm{if} \,\,\vec{c}\neq \vec{d},\eea  
where here and in the following we will use the notation $\vec{c}\cdot\vec{d}=\cos (\gamma)$. 
Note that if both measurements have the same (projective) effects, i.e. $\vec{c}=\vec{d}$, $\BB_{1}$ is independent of $\vrho_{0}$ and hence we do not have to specify $\vrho_{0}$ in this case. 
Analogusly, $\BB_{1}$ is maximized by choosing 
\bea 
\vec{\alpha}_{1}= \frac{\vec{d}-\vec{c}}{\sqrt{2-2\cos (\gamma )} }\,\,\mathrm{if}\,\, \vec{c}\neq \vec{d}.
\eea  
As before, $\BB_{1}$ is independent of $\vrho_{1}$ in case $\vec{d}=\vec{c}$.
Inserting the optimal choice of $\vrho_{0}$ and $\vrho_{1}$ in Eq.(\ref{eqf0}) and using the notation 
$X=2+\sqrt{2-2\cos (\gamma )}$ we obtain
\bea
\frac{X}{4}[2+ (\vec{c}+\vec{d})\cdot\vec{\alpha}_{\rm in}].
\eea
Hence, the optimal choice of $\vrho_{\rm in}$ is given by 
\begin{align}\vec{\alpha}_{\rm in}=\frac{\vec{c}+\vec{d}}{\sqrt{2+2\cos (\gamma)}}\,\, \textrm{if}\,\, \vec{c}\neq-\vec{d}.
\end{align}
Similarly to before, we do not have to specify the input state if $\vec{c}=-\vec{d}$.
Note that the optimal input and post-measurement states are all pure, i.e.  $|\vec{\alpha}_i|=1$ for $i\in\{{\rm in},0,1\}$ and we obtain for these choice of states that $\BB_{1}$ is equal to
\bea\frac{X}{4}[2 +\sqrt{2+2\cos (\gamma)}].
\eea
Note further that this expression only depends on $\cos (\gamma)$.
Considering now the the critical points and the boundary points with respect to $\cos (\gamma)$ one obtains that $\cos (\gamma)\in \{0,1,-1\}$. For $\cos (\gamma)\in \{1,-1\}$ it holds that $\BB_{1}=2$ and for $\cos (\gamma)=0$ we have that  $\BB_{1}=3/2+\sqrt{2}<3$. Hence, $\BB_{1}$ is upper bounded by $3$. As can be easily seen $\BB_{1}=3$ can be attained via the following protocol. We choose $\rho_{\rm in}=\ketbra{0}$, $\rho_{0}=\ketbra{1}$, $\rho_{1}=\ketbra{0}$, $\EE_{0|0}=\openone$ and $\EE_{0|1}=\ketbra{0}$.  This implies that this bound is tight.$\hfill \Box$

\subsubsection{The extreme point $e_2$ and its associated temporal inequality}
It can be easily seen that the extreme point $e_2$
\begin{equation}
e_2: \; p(01|00) = p(01|11) = p(00|01) = p(00|10) = 1
\end{equation}
cannot be reached with measurements on a qubit. In order to do so note that $p(01|00) = p(01|11)=1$ implies that $\MM_0$ and $\MM_1$ are the same non-trivial measurements and have
projective effects. Moreover, the initial state has to be flipped after measuring $\MM_0$ or $\MM_1$. However, this contradicts $p(00|10) = 1$.

Using an analogous argumentation as for the vertex $e_1$ one can analytically show that $\BB_2 =p(01|00) + p(01|11) +p(00|01) + p(00|10) \leq 3.5$. Moreover, it can be shown that the maximum of $\BB_2$ is either given by $3$ or is attained if one of the measurements has projective effects and for the other measurement the effect for outcome $1$ is proportional to a projector. Determining the optimal initial and post-measurement states as before one obtains for this scenario an expression, which depends on solely two parameters. Numerical maximization of this strongly suggests that the maximum of $\BB_2$ is given by $3$. Note that $\BB_2=3$ can be attained by choosing  $\rho_{\rm in}=\ketbra{0}$, $\rho_{0}=\ketbra{0}$, $\rho_{1}=\ketbra{1}$, $\EE_{0|0}=\openone$ and $\EE_{0|1}=\ketbra{0}$.

\subsubsection{The extreme point $e_3$ and its associated temporal inequality}
In this subsection we show that  for measurements on a single qubit $\BB_3=p(01|00)+p(00|11)+p(01|01) + p(01|10)\leq \mathcal{C}_3 \approx 3.186$ and that the bound is attained for measurements, $\mathcal{M}_1$ and $\mathcal{M}_0$, whose effects are projective. 

\noindent{\it Proof.} We will first show that for all initial states and post-measurement states the maximal value of $\BB_3$ is either smaller or equal to 3 or is obtained if all effects of both measurement settings are projectors. However, as we will show the maximum of $\BB_3$ exceeds $3$. We will, then, identify the optimal initial and post-measurement states for such measurements. Using these results $\BB_3$ depends solely on one remaining parameters, the angle between the measurement directions of the effects of $\mathcal{M}_0$ and $\mathcal{M}_1$. With this the maximum of $\BB_3$ can be evaluated by determining the zeros of a polynomial of degree ten which yields to a maximal value of $\BB_3$ given by approximately  $3.186$. 

As in the proof of the upper bound on $\BB_1$, the effects for $\mathcal{M}_0$ ($\mathcal{M}_1$) corresponding to outcome $r\in \{0,1\}$ will be denoted by $\EE_{r|0}$ ($\EE_{r|1}$) respectively and we will first use the the following decomposition for these effects,
\begin{align}\label{effapp211}
&\EE_{0|0}=a_0(\openone + b_0\,\vec{c}\cdot\vec{\sigma}),\\\label{effapp221}
&\EE_{1|0}=\openone-\EE_{0|0},\\\label{effapp231}
&\EE_{0|1}=a_1(\openone + b_1\,\vec{d}\cdot\vec{\sigma}),\\\label{effapp241}
&\EE_{1|1}=\openone-\EE_{0|1},
\end{align}
where $\vec{c},\vec{d}\in\mathbb{R}^3$, $|\vec{c}|=|\vec{d}|=1$, $\vec{\sigma}=(\sigma_1,\sigma_2,\sigma_3)$ with $\sigma_i$ being the Pauli matrices and we choose without loss of generality $b_s\geq 0$  and therefore $0\leq a_s\leq \frac{1}{1+b_s}$ and $b_s\leq 1$ for $s\in\{0,1\}$.
First note that due to the AoT constraints we have that $p(ab|xy)=p(a|x)p(b|axy)$ and hence
\begin{align}\nonumber
\BB_3&=p(0|0)[p(1|000)+p(1|001)]\\&+p(0|1)[p(1|010)+p(0|011)].
\end{align}
We will denote in the following by $\vrho_{\rm in}$ the initial state and by $\vrho_{0}$ ($\vrho_{1}$) the post-measurement states given that measurement $\mathcal{M}_0$ ($\mathcal{M}_1$) has been performed at time $t_1$ and outcome $0$ has been obtained respectively.  Moreover, we will use the notation 
\bea\label{eqrho1}
\vrho_j=\frac{1}{2}(\openone+\vec{\alpha}_{j}\cdot\vec{\sigma})
\eea
for $j\in\{{\rm in}, 0, 1\}$ where $\vec{\alpha}_{j}\in\mathbb{R}^3$ and $|\vec{\alpha}_{j}|\leq 1$. 
Using the decomposition for the effects in Eqs.(\ref{effapp211})-(\ref{effapp241})  we have that 
\begin{align}
&p(0|0)=a_0\,(1+b_0\vec{c}\cdot\vec{\alpha}_{\rm in}),\\
&p(0|1)=a_1\,(1+b_1\vec{d}\cdot\vec{\alpha}_{\rm in}),\\\
&p(1|000)=1-a_0\,(1+b_0\vec{c}\cdot\vec{\alpha}_{0}),\\
&p(1|001)=1-a_1\,(1+b_1\vec{d}\cdot\vec{\alpha}_{0}),\\
&p(1|010)=1-a_0\,(1+b_0\vec{c}\cdot\vec{\alpha}_{1}),\\
&p(0|011)=a_1\,(1+b_1\vec{d}\cdot\vec{\alpha}_{1}).
\end{align}
We will next show that for any $\vrho_{\rm in}$, $\vrho_{0}$, $\vrho_{1}$ the maximum of $\BB_3$ is attained for $a_s=\frac{1}{1+b_s}$ for $s\in\{0,1\}$ or is smaller or equal to 3. In order to do so we consider $\BB_3$ first as a function of $a_0$ (all other parameters are fixed but arbitrary) and calculate its critical points. The derivate of $\BB_3$ with respect to $a_0$ is given by
\begin{align}\nonumber
\frac{d\BB_3}{d a_0}&=[p(1|000)+p(1|001)]\,(1+b_0\vec{c}\cdot\vec{\alpha}_{\rm in}) \\&-p(0|0)\,(1+b_0\vec{c}\cdot\vec{\alpha}_{0})-p(0|1)\,(1+b_0\vec{c}\cdot\vec{\alpha}_{1}).
\end{align}
Multiplying this equation by $a_0$ one obtains that for the critical points it has to hold that 
\begin{align}\nonumber
&[p(1|000)+p(1|001)]p(0|0) +p(0|1)p(1|010)\\&=p(0|0)[1-p(1|000)]+p(0|1)\leq 2,
\end{align}
which implies that  $\BB_3$ at the points where the derivative vanishes cannot exceed 3. However, one can easily verify that $\BB_3$ can reach a value of $3.186$ for $a_0=a_1=1/2$, $b_0=b_1=1$, $\cos (\gamma)=\vec{c}\cdot\vec{d}=0.756$ and $\vrho_{\rm in},$ $\vrho_{0},$ $\vrho_{1}$ as given in Eqs.(\ref{rho00opt1}),(\ref{rho11opt1}) and (\ref{rhoinopt1}). Hence, the maximum has to be attained at the boundary of the domain for $a_0$, i.e. it has to hold for the maximum of $\BB_3$
that either $a_0=0$ or $a_0=\frac{1}{1+b_0}$.
It is straightforward to see that for $a_0=0$ it holds that $\BB_3\leq 2$ and therefore the maximum is attained for $a_0=\frac{1}{1+b_0}$.
Analogously, we consider $\BB_3$ as a function of $a_1$ (with all other parameters fixed but arbitrary) and compute the corresponding critical points. The derivative is given by 
\begin{align}\nonumber
\frac{d\BB_3}{d a_1}&=[p(1|010)+p(0|011)]\,(1+b_1\vec{d}\cdot\vec{\alpha}_{\rm in})\\ &-p(0|0)\,(1+b_1\vec{d}\cdot\vec{\alpha}_{0})+p(0|1)\,(1+b_1\vec{d}\cdot\vec{\alpha}_{1})
\end{align}
and therefore it has to hold for any critical point that 
\begin{align}\nonumber
&p(0|0)p(1|001)+[p(1|010)+p(0|011)]p(0|1)
\\&=p(0|0)-p(0|1)p(0|011)\leq 1.
\end{align}
Note that this implies that at a critical points $\BB_3\leq 2$.
At the boundary point given by $a_1=0$ one obtains that  $\BB_3\leq 2$. Hence the maximum of $\BB_3$ is attained for $a_1=\frac{1}{1+b_1}$.
Using that $a_s=\frac{1}{1+b_s}$ the effects of the measurements in Eqs.(\ref{effapp211})-(\ref{effapp241})  can equivalently (substituting $b_0=\frac{p}{2-p}$ and $b_1=\frac{q}{2-q}$) written as 
\begin{align}\label{effapp11}
&\EE_{0|0}=\frac{1}{2}[(2-p) \openone + p \,\vec{c}\cdot\vec{\sigma}],\\\label{effapp21}
&\EE_{1|0}=\frac{p}{2}(\openone -\, \vec{c}\cdot\vec{\sigma}),\\\label{effapp31}
&\EE_{0|1}=\frac{1}{2}[(2-q) \openone + q \,\vec{d}\cdot\vec{\sigma}],\\\label{effapp41}
&\EE_{1|1}=\frac{q}{2}(\openone -\vec{d}\cdot\vec{\sigma}),
\end{align} where $0\leq p\leq 1$ and $0\leq q\leq 1$. 
Considering $\BB_3$ as a function of $q$ one obtains for its derivative
\begin{align}\nonumber 
\frac{d\BB_3}{d q}&=-[p(1|010)+p(0|011)]\,\frac{1}{2}(1-\vec{d}\cdot\vec{\alpha}_{\rm in}) \\&+p(0|0)\,\frac{1}{2}(1-\vec{d}\cdot\vec{\alpha}_{0})-p(0|1)\,\frac{1}{2}(1-\vec{d}\cdot\vec{\alpha}_{1}).
\end{align}
Hence, at the critical points we have that 
\begin{align}\nonumber
0&=p(0|0)p(1|001)+[p(1|010)+p(0|011)][p(0|1)-1]
\\\nonumber&+p(0|1)[p(0|011)-1]
\\\nonumber&\geq p(0|0)p(1|001)+2[p(0|1)-1]+p(0|1)[p(0|011)-1]
\\\nonumber&=p(0|0)p(1|001)+p(0|1)[p(0|011)+1]-2
\\&\geq p(0|0)p(1|001)+p(0|1)[p(0|011)+p(1|010)]-2.
\end{align}
Here we used for the first inequality that $[p(1|010)+p(0|011)]\leq 2$ and $[p(0|1)-1]\leq 0$ and for the second inequality that $p(0|1)\geq 0$ and $p(1|010)\leq 1$.
Note that this implies that $\BB_3\leq 3$ at the points where this derivative vanishes.  It is straightforward to see that for the boundary point $q=0$ one obtains that $\BB_3\leq 3$ and therefore the maximum of $\BB_3$ is attained for the other boundary point, $q=1$.
We will next consider $\BB_3$ as a function of $p$ and compute its critical points
\begin{align}\nonumber
\frac{d\BB_3}{d p}&=-[p(1|000)+p(1|001)]\,\frac{1}{2}(1-\vec{c}\cdot\vec{\alpha}_{\rm in}) \\ &+p(0|0)\,\frac{1}{2}(1-\vec{c}\cdot\vec{\alpha}_{0})+p(0|1)\,\frac{1}{2}(1-\vec{c}\cdot\vec{\alpha}_{1}).
\end{align}
With this we have that at the critical points it holds that 
\bea\nonumber
&p(0|1)p(1|010)+[p(1|000)+p(1|001)]p(0|0)\\&=p(1|000)+p(1|001)-p(0|0)p(1|000)\leq 2
\eea
and therefore $\BB_3\leq 3$ at the critical points. It is easy to see that for $p=0$ it holds that $\BB_3\leq 2$. Hence, the maximum of $\BB_3$ is attained at the boundary $p=1$. Note that, in summary, we have shown that the optimal measurements have projective effects.
Using Eq. (\ref{eqrho1}), as well as Eqs. (\ref{effapp11})-(\ref{effapp41}) for $q=p=1$ we have that $\BB_3$ is of the following form
\begin{align}\nonumber\BB_3&=\frac{1}{4}(1+ \vec{c}\cdot\vec{\alpha}_{\rm in})[2-(\vec{d}+\vec{c})\cdot\vec{\alpha}_{0}]\\\label{eqfB3}&+\frac{1}{4}(1+\vec{d}\cdot\vec{\alpha}_{\rm in})[2+(\vec{d}-\vec{c})\cdot\vec{\alpha}_{1}].
\end{align}
As $\frac{1}{2}(1+ \vec{c}\cdot\vec{\alpha}_{\rm in})=p(0|0)\geq 0$ it has to hold for the maximum of $\BB_3$  that
\bea \label{rho00opt1}
\vec{\alpha}_{0}= \frac{-(\vec{d}+\vec{c})}{\sqrt{2+2\cos (\gamma )}}\,\, \mathrm{if} \,\,\vec{d}\neq -\vec{c},\eea  
where here and in the following $\cos (\gamma)=\vec{c}\cdot\vec{d}$. 
Note that if $\vec{d}=-\vec{c}$ then $\BB_3$ is independent of $\vrho_{0}$ and hence we do not have to specify $\vrho_{0}$ in this case. 
Analogously, $\BB_3$ is maximized by choosing 
\bea \label{rho11opt1}
\vec{\alpha}_{1}= \frac{(\vec{d}-\vec{c})}{\sqrt{2-2\cos (\gamma )} }\,\,\mathrm{if}\,\, \vec{d}\neq\vec{c}.
\eea  
Similarly to before, $\BB_3$ is independent of $\vrho_{1}$ if $\vec{d}=\vec{c}$.
Inserting the optimal choice of $\vrho_{0}$ and $\vrho_{1}$ in Eq.(\ref{eqfB3}) and using the notation 
$X_0=2+\sqrt{2+2\cos (\gamma)}$ and $X_1=2+\sqrt{2-2\cos (\gamma)}$ we obtain
\bea
\frac{1}{4}[X_0+ X_1+(X_0\vec{c}+X_1\vec{d})\cdot\vec{\alpha}_{\rm in}].
\eea
Hence, the optimal choice of $\vrho_{\rm in}$ is given by 
\begin{align}\label{rhoinopt1}
\vec{\alpha}_{\rm in}=\frac{X_0\vec{c}+X_1\vec{d}}{\sqrt{X_0^2+X_1^2+2X_0X_1\cos (\gamma)}}\\\notag
\,\, \textrm{if}\,\, X_0\vec{c}+X_1\vec{d}\neq 0.
\end{align}
Analogously to before, we do not have to specify the input state if $X_0\vec{c}+X_1\vec{d}=0$.
Note that the optimal input and post-measurement states are all pure, i.e.  $|\vec{\alpha}_i|=1$ for $i\in\{{\rm in},0,1\}$ and we obtain for these choice of states that $\BB_3$ is equal to
\begin{align}
&\frac{1}{4}[X_0+X_1
+\sqrt{X_0^2+ X_1^2+2X_0X_1\cos (\gamma)}].
\end{align}
Note that this equation only depends on a single parameter namely the angle, $\gamma$. For the points at the boundary and the points for which the derivative with respect to $\cos(\gamma)$ is not defined which all are given by $\cos(\gamma)\in\{1,-1\}$ one can show that $\BB_3\leq 3$.  Hence, in this case the maximum of $\BB_3$  can be determined by finding the point were the derivative with respect to $\cos(\gamma)$ vanishes. For this one has to solve the polynomial equation \begin{align}\notag
0=&1 -x (42 - x (-531 -    4 x (380 -         x \\\notag
&\cdot (-24 -              x (-762 -                x (481 - 8 x \\\label{eq:polynom1}
&\cdot (19 - 4 x (-3 + 2 (1 + x) x)))))))),
\end{align}
with $x = \cos \gamma $, and determine the solution for which $\frac{d\BB_3}{d\cos(\gamma)}$ is indeed $0$, which yields approximately $3.186$. Note that the derivative  $\frac{d\BB_3}{d\cos(\gamma)}=0$ was squared multiple times in order to arrive at Eq. (\ref{eq:polynom1}) which created additional roots that are not solutions of the original equation. $\hfill \Box$

\subsubsection{The extreme point $e_4$ and its associated temporal inequality}
It is straightforward to see that the same argumentation that has been presented in order to show that the vertex $e_2$ cannot be reached on a qubit applies also to the extreme point $e_4$
\begin{equation}
e_4: \; p(01|00) = p(01|11) = p(01|01) = p(00|10) = 1.
\end{equation}
Using analogous methods to before it can be shown that for a qubit the value of $\BB_4=p(01|00) + p(01|11) + p(01|01) + p(00|10)$ is either smaller or equal to 3 or is obtained if the effect of $\mathcal{M}_0$ for the outcome 1 has rank $1$ and the effects of $\mathcal{M}_1$ are projectors. However, it can be easily verified that $\BB_4$ exceeds $3$. Identifying the optimal initial and post-measurement states for such measurements, which are all pure,  one obtains the following expression for $\BB_4$ 
\begin{align}\label{eqoptstatese4}
&\frac{1}{4}[(2-p) X_0+X_1
+\sqrt{p^2 X_0^2+ X_1^2+2pX_0X_1\cos (\gamma)}],
\end{align}
where the effects of the measurements have been parametrized as in Eqs. (\ref{effapp11})- (\ref{effapp41}) with $q=1$, $\cos (\gamma)=\vec{c}\cdot\vec{d}$, $X_0=1+p+\sqrt{p^2+1+2p\cos (\gamma)}$ and $X_1=3-p+\sqrt{p^2+1-2p\cos (\gamma)}$.
Note that this expression depends solely on the remaining 2 parameters of the effects of $\mathcal{M}_0$ and $\mathcal{M}_1$.
Performing a numerical optimization of this expression strongly suggests that the maximum of  $\BB_4$ is approximately $3.186$ and is attained for measurements which have projective effects. Note that if one restricts $\mathcal{M}_0$, $\mathcal{M}_1$ to measurements for which all effects have rank 1 then $\BB_4$ is equivalent to $\BB_3$. Moreover, one can analytically show that $\BB_4\leq 2+\sqrt{2}$. In order to do so note that due to  $pX_0, X_1\geq 0$ and  $a^2+b^2+2ab\cos (\beta)\leq (a+b)^2$ $\forall a,b\geq 0$ and $\beta\in\mathbb{R}$  we have that the maximum of $\BB_4$  is upper bounded as follows
\begin{align}\notag
&\frac{1}{4}[(2-p) X_0+X_1
+\sqrt{p^2 X_0^2+ X_1^2+2pX_0X_1\cos (\gamma)}]\\
&\leq \frac{1}{2}(X_0+ X_1)\\\nonumber
&= \frac{1}{2}(4+\sqrt{p^2+1-2p\cos (\gamma)}+\sqrt{p^2+1+2p\cos (\gamma)}).
\end{align} Considering this expression as a function of $\cos (\gamma)$ and computing its critical points it is then straightforward to see that this upper bound for $\BB_4$ does not exceed $2+\sqrt{2}$.

We also considered a different parametrization to evaluate the maximum of $\BB_4$ numerically. First, we expressed the temporal Bell operator $\BB_4$ in terms of the Kraus operators of the measurements and calculated the maximal expectation value with a pure state, i.e. we maximized
\begin{align}\nonumber
\BB_4 &= \bra{\psi}2K_{00}^\dagger K_{00} +K_{11}^\dagger K_{11}  + K_{11}^\dagger\left(K_{00}^\dagger K_{00}\right)K_{11}\\\nonumber  &- K_{00}^\dagger \left(K_{11}^\dagger K_{11} + K_{00}^\dagger K_{00}\right)K_{00}\\  &-K_{11}^\dagger \left(K_{11}^\dagger K_{11}\right)K_{11}\ket{\psi},
\end{align}
with $K_{00}^\dagger K_{00} = \EE_{0|0}$ and $K_{11}^\dagger K_{11} = \EE_{1|1}$ under the constraint that $\braket{\psi} {\psi} = 1$ and $0\leq \EE_{i|i} \leq \mathds{1}$ . Note that due to the fact that the maximum is not only attained for pure initial states but also the post-measurement states should be pure it is sufficient to consider a single Kraus operator per effect. Using this parametrization we numerically evaluated the maximum of $\BB_4$ for measurements on a single qubit to be approximately $3.186$. 

\section{Appendix D: Lower bounds on $\epsilon$ for $(2+\epsilon)$-dimensional systems}
Let us first recall our definition of a $(d+\epsilon)$-dimensional system. A system, i.e. an initial state $\vrho_{\rm in}$ and a set of measurements with instruments $\{\II_{a|x}\}_a$, has dimension $d+\epsilon$ (with $\epsilon\geq 0$) if there exists a projector on a d-dimensional subspace, $P_d$, such that $\lVert P_d\vrho_{\rm in}P_d-\vrho_{\rm in} \rVert_{\rm tr}\leq \epsilon$ and for all $a, x$ and quantum states $\rho$ it holds that $\lVert P_d\II_{a|x}(\rho)P_d-\II_{a|x}(\rho) \rVert_{\rm tr}\leq \epsilon$.  Hence, a $(d+\epsilon)$-dimensional systems is a system for which the initial states as well as all possible post-measurement states of the instruments deviate only by $\epsilon$ from the same d-dimensional subspace.

In the following we will establish lower bounds on $\epsilon$ for $(2+\epsilon)$-dimensional systems. In particular, we will provide lower bounds that are determined each by the expectation value $\mathcal{B}_i$ which has been defined in the main text (see also Appendix C). Hence, these lower bounds can be accessed in an experiment. Moreover, this implies that a value of $\mathcal{B}_i$ that is larger than the bound $C_i$ for measurements on a qubit does not only allow to conclude that the measurements are performed on a qutrit but also provides some way to quantify how close the system is to a qubit.

In order to establish these lower bounds we consider the conditional probability distribution $p(ab|xy)$ for a $(2+\epsilon)$-dimensional system. In the following we will denote $\tilde{\II}_{a|x}(\vrho)\equiv P_2\II_{a|x}(\rho)P_2$ and $\tilde{\vrho}_{\rm in}\equiv P_2\vrho_{\rm in}P_2$, where $ P_2$ is the projector on the two-dimensional subspace from which the system deviates by $\epsilon$. Moreover, we will use that for all hermitian operators $M$ it holds that $\tr (M) \leq \lVert M \rVert_{\rm tr} $ and that the completely positive maps $\II_{a|x}$ are trace non-increasing, i.e. for each quantum state $\rho$ there exists a quantum state $\sigma_\rho^{a,x}$ and a probability $p_\rho^{a,x}$ such that $\II_{a|x}(\rho)=p_\rho^{a,x}\, \sigma_\rho^{a,x}$. 
Hence, we have that 
\begin{align}\nonumber
p(ab|xy)=&\tr \{\II_{b|y}[\II_{a|x}(\vrho_{\rm in})]\}\\\nonumber
=&\tr \{(\II_{b|y}-\tilde{\II}_{b|y})[\II_{a|x}(\vrho_{\rm in})]\}\\\nonumber &+\tr \{\tilde{\II}_{b|y}[\II_{a|x}(\vrho_{\rm in})]\}\\
=&p_{\vrho_{\rm in}}^{a,x}\tr \{(\II_{b|y}-\tilde{\II}_{b|y})(\sigma^{a,x}_{\vrho_{\rm in}})\}\\\nonumber&+tr \{\tilde{\II}_{b|y}[\II_{a|x}(\vrho_{\rm in})]\}\\
\leq& \tr \{\tilde{\II}_{b|y}[\II_{a|x}(\vrho_{\rm in})]\}+\epsilon\\
=&\tr \{\tilde{\II}_{b|y}[(\II_{a|x}-\tilde{\II}_{a|x})(\vrho_{\rm in})]\}\\\nonumber&+\tr \{\tilde{\II}_{b|y}[\tilde{\II}_{a|x}(\vrho_{\rm in})]\}+\epsilon\\
\leq& \tr \{(\II_{a|x}-\tilde{\II}_{a|x})(\vrho_{\rm in})]\}\\\nonumber&+\tr \{\tilde{\II}_{b|y}[\tilde{\II}_{a|x}(\vrho_{\rm in})]\}+\epsilon\\
\leq& \tr \{\tilde{\II}_{b|y}[\tilde{\II}_{a|x}(\vrho_{\rm in})]\}+2\epsilon\\
=& \tr \{\tilde{\II}_{b|y}[\tilde{\II}_{a|x}(\tilde{\vrho}_{\rm in})]\}+2\epsilon\\\nonumber&+\tr \{\tilde{\II}_{b|y}[\tilde{\II}_{a|x}(\vrho_{\rm in}-\tilde{\vrho}_{\rm in})]\}\\
\leq& \tr \{\tilde{\II}_{b|y}[\tilde{\II}_{a|x}(\tilde{\vrho}_{\rm in})]\}+3\epsilon.
\end{align}
Note that the maximum of $\sum_{a,x,b,y} q_{a,x,b,y}^i \tr \{\tilde{\II}_{b|y}[\tilde{\II}_{a|x}(\tilde{\vrho}_{\rm in})]\}$ with $q_{a,x,b,y}^i=p(ab|xy)$ of the extreme point $e_i$ is upper bounded by $C_i$ as for $x=0, 1$ $\sum_a \tilde{\II}_{a|x}$  is a trace-nonincreasing map  (but not necessarily necessarily trace-preserving). With this one obtains that for a $(2+\epsilon)$-dimensional system
\bea
\mathcal{B}_i\leq C_i+12\epsilon,
\eea
where $C_i$ denotes as before the bound obtained for measurements on a qubit. This provides the following lower bound on $\epsilon$
\bea
\epsilon \geq \frac{\mathcal{B}_i-C_i}{12},
\eea
which can be evaluated in an experiment by determining $\mathcal{B}_i$. Note, however, that the maximum possible value of the lower bound is given by by  $\frac{4-C_i}{12}$ and hence larger values for $\epsilon$ cannot be certified by using this scheme.


\begin{thebibliography}{99}
\bibitem{footnote} Contrary to the case of POVMs, which can be directly
seen as projective measurement on a higher dimensional space (Neumark
dilation), quantum instruments require a unitary transformation and a
projective measurement (Stinespring dilation), e.g., coupling to an
ancilla and projective measurement on the ancilla. This is the reason
why methods for bounding temporal correlations for projective
measurements do not generalize straightforwardly to arbitrary instruments.
\bibitem{footnote2} At each time step the same relabeling is performed. All possible relabelings that fulfill this constraint are allowed.
\bibitem{footnote3} The bounds provided in Theorem 2 hold true even if one allows for arbitrary (not necessarily Markovian) state transformation between the measurement. This is due to the fact that for sequences of length two any such state transformation only leads to a modification of the instruments.
\end{thebibliography}

\end{document}